\documentclass[final,5p,times,twocolumn,numbers,sort&compress]{elsarticle}
\usepackage{amsmath,amssymb}
\usepackage{graphicx}
\usepackage{subcaption}
\usepackage{booktabs}
\usepackage{multirow}
\usepackage{siunitx}
\usepackage{doi}
\usepackage{hyperref}
\hypersetup{colorlinks=true,linkcolor=blue,citecolor=blue}

\journal{Physics Letters B}

\begin{document}
	
	\begin{frontmatter}
		
		\title{Minimal Extensions of the \texorpdfstring{\(\alpha\)}{alpha}-Starobinsky Model: Reconciling ACT DR6 and Reheating Constraints}
		
		\author[1]{Nehla Shobcha}
		\author[2]{Norma Sidik Risdianto}
		\author[1]{Romy Hanang Setya Budhi \corref{cor1}}
		
		\cortext[cor1]{Corresponding author.}
		\ead{romyhanang@ugm.ac.id}
		
		\affiliation[1]{organization={Department of Physics, Faculty of Mathematics and Natural Sciences, Universitas Gadjah Mada},
			city={Yogyakarta},
			postcode={55281},
			country={Indonesia}}
		\affiliation[2]{organization={Department of Physics Education, Universitas Islam Negeri Sunan Kalijaga},
			city={Yogyakarta},
			postcode={55281},
			country={Indonesia}}
		
		\begin{abstract}
			The latest combined data from the Atacama Cosmology Telescope (ACT) DR6, Planck, and DESI yields a scalar spectral index $n_s = 0.9743 \pm 0.0034$, which lies approximately $2\sigma$ above the prediction of the standard $\alpha$-Starobinsky inflation model. To address this tension, we propose two minimal extensions that preserve the model's plateau structure and attractor properties: a multiplicative exponential modification and an additive polynomial deformation, both governed by a single small perturbative parameter $\delta>0$. We analytically derive the slow-roll parameters and inflationary observables up to second order in $\delta$ and integrate them with reheating dynamics via the consistency equation. It is shown that the $\delta$ term effectively shifts $n_s$ into the $1\sigma$ confidence region of the joint P-ACT-LB-BK18 dataset without violating the tensor-to-scalar ratio bound ($r < 0.038$). The viable parameter space at $1\sigma$ requires $\alpha \lesssim 35$ with $\delta \sim \mathcal{O}(10^{-2})$ for the exponential model, while the additive model requires $\delta \sim \mathcal{O}(10^{-3})$ for $p=1$ and $\delta \sim \mathcal{O}(10^{-4})$ for $p=2$. For the e-folding range $N_k \in [50, 65]$, the relevant reheating equation of state is $0<\omega_{\mathrm{re}}\le1$. All viable scenarios yield a reheating temperature $T_{\mathrm{re}} \sim 10^9$ GeV, which is safely above the Big Bang Nucleosynthesis (BBN) bound and below the gravitino overproduction limit.
		\end{abstract}
		
		
	\end{frontmatter}
	
	\section{Introduction}
	\label{sec:intro}

	While the $\alpha$-Starobinsky model has long been consistent with Planck 2018 data, the latest ACT DR6 release, when combined with Planck, BICEP/Keck, and DESI (P-ACT-LB-BK18), pushes the scalar spectral index to $n_s = 0.9743 \pm 0.0034$ \cite{ACT:2025fju, ACT:2025tim, Adame_2025-1, Adame_2025-2}. This value lies about $2\sigma$ above the standard prediction of $n_s \approx 0.960$ to $0.967$ for the e-fold number at horizon crossing  $N_k \sim 55$, creating significant tension. Several extensions have been proposed to resolve this \cite{Odintsov:2025, Yi:2025, Addazi:2025, Dioguardi:2025, Mohammadi:2025, Drees:2025}, yet the constraints imposed by the post-inflationary reheating phase are rarely taken into account.

	The reheating phase is a critical component that determines the thermal history of the early universe.  The reheating consistency equation relates $N_k$ non-linearly to both the reheating temperature $T_{\mathrm{re}}$ and the effective equation of state $\omega_{\mathrm{re}}$ \cite{Liddle:2003as, German:2024rmn, Bassett:2005xm, Podolsky:2005bw, Allahverdi:2010xz, Kofman:1994rk, German:2025xxx}. This dynamics is subject to strict bounds, including a lower limit from Big Bang Nucleosynthesis (BBN, $T_{\mathrm{re}} \gtrsim 10$ MeV) and an upper limit that prevents gravitino overproduction or excessive modification of relativistic degrees of freedom via primordial gravitational waves \cite{Haque:2021, Mohammadi:2025, Kohri:2005wn, Khlopov:1984pf, Bolz:2000fu}. Reheating constraints are essential for breaking the degeneracy among inflationary models that predict nearly identical $\{n_s, r\}$ values with weak parameter sensitivity \cite{Mishra:2021,Bahari:2025ttf, Drees:2025}. Any extension of the $\alpha$-Starobinsky model can therefore be distinguished through its reheating predictions.

	In this paper, we propose two minimal extensions of the $\alpha$-Starobinsky model that preserve its plateau structure and attractor properties, specifically  a multiplicative exponential modification and an additive polynomial deformation. We analytically derive the slow-roll parameters and inflationary observables up to second order in the perturbative parameter $\delta$, examine the reheating constraints, and map the allowed parameter space for $\alpha$, $\delta$, and $\omega_{\mathrm{re}}$. The paper is organized as follows. Section 2 summarizes the formalism of inflation and reheating. Section 3 discusses the exponential and additive extensions in detail. Section 4 presents our conclusions.

	\section{Inflation and post-inflationary reheating}
	
	\subsection{Inflation and its observables}
	
	Cosmic inflation is a phase of extremely rapid expansion in the early universe that resolves the horizon and flatness problems and provides a mechanism for generating primordial fluctuations \cite{Martin:2013tda, Guth:1981, Starobinsky:1980te}. In the single-field framework, the inflaton $\phi$ with potential $V(\phi)$ satisfies the Klein-Gordon  $\ddot{\phi} + 3H\dot{\phi} + V'(\phi) = 0,$ and Friedmann equations in an FLRW background $H^2 = \frac{1}{3M_{\mathrm{Pl}}^2}\left(\frac{1}{2}\dot{\phi}^2 + V(\phi)\right),$
	where $H=\dot{a}/a$ is the Hubble parameter and $M_{\mathrm{Pl}}\approx2.44\times10^{18}$ GeV is the reduced Planck mass.
	
	During inflation, the field evolves slowly such that the slow-roll approximation holds: $3H\dot{\phi} \approx -V'(\phi)$ and $H^2 \approx V(\phi)/(3M_{\mathrm{Pl}}^2)$. The validity of this approximation is quantified by the smallness of the slow-roll parameters \cite{Martin:2013tda, Guth:1981}:
	\begin{equation}\label{eq:slowroll}
		\epsilon \equiv \frac{M_{\mathrm{Pl}}^2}{2}\left(\frac{V'}{V}\right)^2,\qquad
		\eta \equiv M_{\mathrm{Pl}}^2\frac{V''}{V},\qquad
		\xi_2 \equiv M_{\mathrm{Pl}}^4\frac{V'V'''}{V^2}.
	\end{equation}
	Inflation persists while $\epsilon<1$ and terminates when $\epsilon(\phi_{\mathrm{end}})=1$. The number of e-folds between the horizon crossing of the CMB mode with wavenumber $k$ and the end of inflation is
	\begin{equation}
		N_k \equiv \ln\frac{a_{\mathrm{end}}}{a_k} = \int_{t_k}^{t_{\mathrm{end}}} H\,dt \approx \frac{1}{M_{\mathrm{Pl}}^2}\int_{\phi_{\mathrm{end}}}^{\phi_k}\frac{V}{V'}\,d\phi.
	\end{equation}
	Resolving the horizon problem typically requires $N_k\approx50$ to $60$.
	
	The inflationary observables are determined at horizon crossing by the slow-roll parameters  as
	\begin{align}
		n_s = 1+2\eta-6\epsilon,\quad r = 16\epsilon,\quad n_{sk} = 16\epsilon\eta-24\epsilon^2-2\xi_2, \label{eq:ns_r_nsk}
	\end{align}
	Planck 2018 combined with BICEP/Keck 2018 and BAO data yield \cite{Planck:2020vi,Akrami:2018odb,BICEP:2021xfz}
	\begin{equation}
		n_s = 0.9649 \pm 0.0042\;(68\%\text{ CL}),\quad r < 0.036\;(95\%\text{ CL}).
	\end{equation}
	The latest results from the Atacama Cosmology Telescope (ACT) DR6 \cite{ACT:2025fju,ACT:2025tim}, when combined with Planck 2018, BICEP/Keck 2018 (BK18), lensing data, and DESI (the P-ACT-LB-BK18 dataset), provide a higher spectral index 
	\begin{equation}
		n_s = 0.9743 \pm 0.0034\;(68\%\text{ CL}),
	\end{equation}
	together with a tighter upper bound on the tensor-to-scalar ratio, $r < 0.038$ (95\% CL) \cite{ACT:2025tim, Drees:2025}. This central value of $n_s$ lies about $2\sigma$ above the Planck 2018 result, thereby pressuring single-field inflation models such as $\alpha$-Starobinsky, which generally predict $n_s < 0.97$ for $\alpha \sim \mathcal{O}(1)$ \cite{German:2024rmn}. This tension motivates the minimal extensions developed in the present work.
	
\begin{figure}[!t]
		\centering
		\includegraphics[width=0.8\columnwidth]{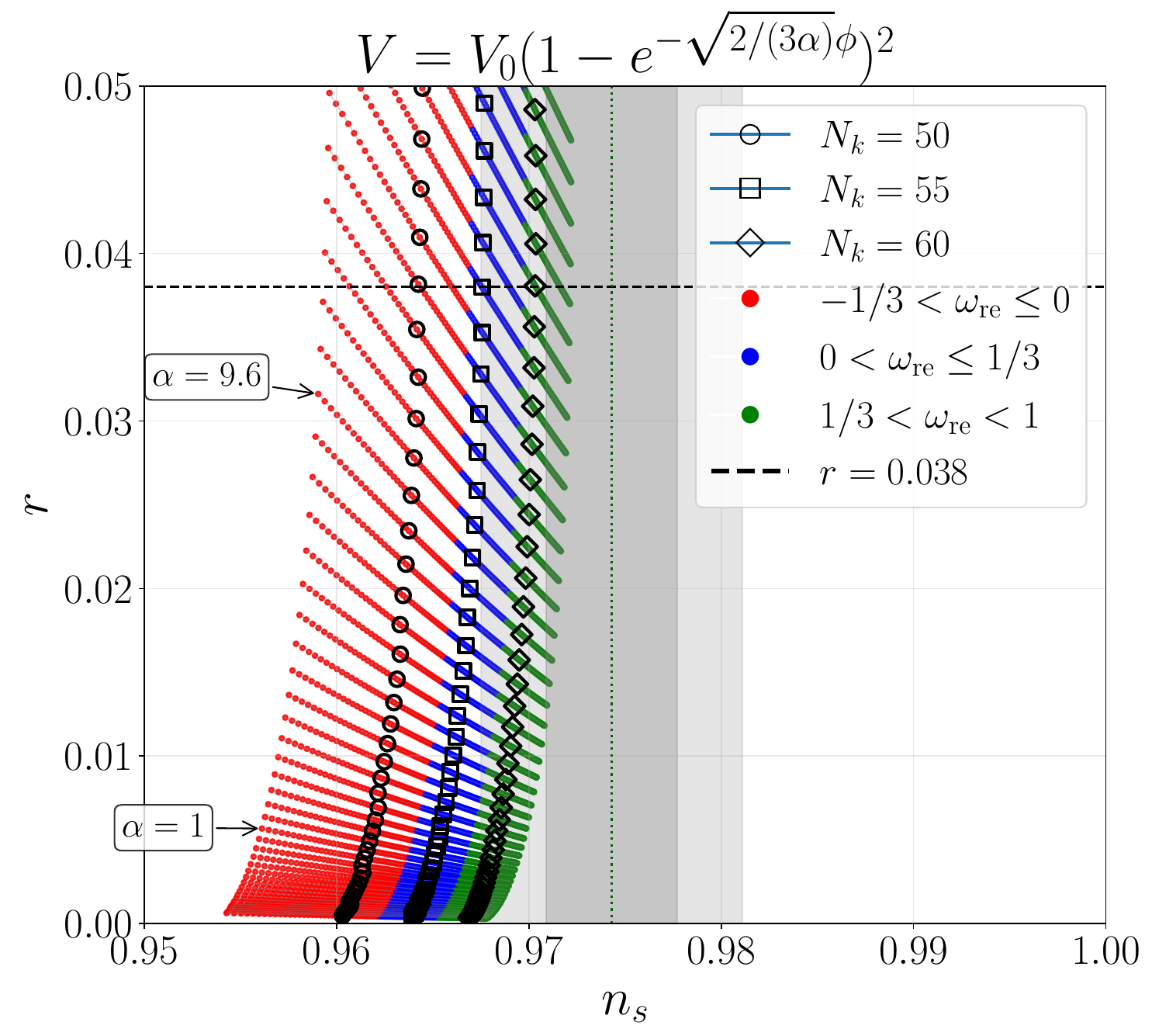}
		\caption{Predictions of the $\alpha$-Starobinsky model in the $n_s$-$r$ plane against the $1\sigma$ (dark gray) and $2\sigma$ (light gray) confidence regions from the P-ACT-LB-BK18 data. The curves are separated according to the equation of state segments $\omega_{\mathrm{re}}$ for each $\alpha$.}
		\label{fig:alpha_starobinsky}
	\end{figure}

	The $\alpha$-Starobinsky model is a one-parameter generalization of the original Starobinsky model \cite{Starobinsky:1980te} belonging to the class of inflationary plateau models \cite{Ellis:2013nxa,Kallosh:2013yoa,Ellis:2015xna}. In terms of the variable $u=\exp~\bigl(-\sqrt{2/(3\alpha)}\,\phi/M_{\mathrm{Pl}}\bigr)$, the potential takes the simple form $V(u)=V_0(1-u)^2$. When $\alpha=1$, the model reduces to the pure Starobinsky form. This model has been extensively tested against CMB data and remains highly competitive \cite{Saini:2025, German:2020iwg}. Inflation ends when $\epsilon(u_{\mathrm{end}})=1$, which yields $u_{\mathrm{end}}=\sqrt{3\alpha}/(2+\sqrt{3\alpha})$, and the number of e-folds from horizon crossing to the end of inflation is
	\begin{equation}
		N_k=\frac{3\alpha}{4}\left(\frac1{u_k}-\frac1{u_{\mathrm{end}}}+\ln\frac{u_{\mathrm{end}}}{u_k}\right).
	\end{equation}
	In the limit $N_k\gg1$, one obtains $u_k\approx3\alpha/(4N_k)$, from which the inflationary observables follow as
	\begin{equation}
		n_s^{(0)}\approx1-\frac{2}{N_k}-\frac{9\alpha}{2N_k^2},\quad
		r^{(0)}\approx\frac{12\alpha}{N_k^2},\quad
		n_{sk}^{(0)}\approx-\frac{2}{N_k^2}-\frac{9\alpha}{2N_k^3}.
	\end{equation}
	For $N_k=55$ and $\alpha=1$, these expressions give $n_s^{(0)}\approx0.962$ and $r\approx0.004$; for $\alpha=10$, $n_s^{(0)}\approx0.967$ and $r\approx0.01$. These predictions remain consistent with Planck 2018 \cite{Akrami:2018odb}. Once reheating dynamics are taken into account, however, $N_k$ is no longer a free parameter but is fixed by the consistency equation \eqref{eq:consistency}.
	Recent studies demonstrate that the $\alpha$-Starobinsky model can still satisfy Planck 2018 constraints over a limited range of $N_k$, yet it cannot reach $n_s\approx0.974$ as required by the ACT DR6 data \cite{German:2024rmn,Drees:2025}. The numerical results of this model are illustrated in Fig.~\ref{fig:alpha_starobinsky}. This shortcoming provides the primary motivation for the model extensions discussed in the next section.
	
	\subsection{Reheating and the consistency relation}
	
	After inflation ends ($\epsilon=1$), the inflaton oscillates around the minimum of its potential, dominating the universe as an effective fluid with equation of state $\omega_{\mathrm{re}}$, bounded within $-1/3<\omega_{\mathrm{re}}<1$ \cite{Cook:2015, Dai:2014}. The stored energy is transferred to radiation through reheating, which sets the temperature $T_{\mathrm{re}}$, duration $N_{\mathrm{re}}$, and leaves imprints on nucleosynthesis and the CMB \cite{Podolsky:2005bw,Turner:1983he, Bassett:2005xm, Allahverdi:2010xz,Amin:2014eta}. Reheating ends when $H(t_{\mathrm{re}})\simeq\Gamma_\phi$, equilibrating the inflaton and radiation energy densities \cite{Kofman:1994rk,Podolsky:2005bw, Amin:2014eta}:
	\begin{equation}
		\rho_{\mathrm{re}} \approx 3\Gamma_\phi^2 M_{\mathrm{Pl}}^2 = \frac{\pi^2}{30}g_{\mathrm{re}}T_{\mathrm{re}}^4,
	\end{equation}
	with $g_{\mathrm{re}}\simeq 106.75$ the effective relativistic degrees of freedom. The reheating duration, for constant $\omega_{\mathrm{re}}$, is
	\begin{equation}
		N_{\mathrm{re}} \equiv \ln\!\left(\frac{a_{\mathrm{re}}}{a_{\mathrm{end}}}\right) = \frac{1}{3(1+\omega_{\mathrm{re}})}\ln\!\left(\frac{\rho_{\mathrm{end}}}{\rho_{\mathrm{re}}}\right),\qquad
		\rho_{\mathrm{end}} \approx \frac{3}{2}V(\phi_{\mathrm{end}}),
	\end{equation}
	where $a_{\mathrm{end}}$ and $a_{\mathrm{re}}$ are the scale factors at the end of inflation and reheating. The subsequent radiation era has $\omega=1/3$, and the number of e-folds from reheating to matter--radiation equality is
	\begin{equation}
		N_{\mathrm{rd}} \equiv \ln\!\left(\frac{a_{\mathrm{eq}}}{a_{\mathrm{re}}}\right) = \ln\!\left(\frac{a_{\mathrm{eq}}T_{\mathrm{re}}}{(43/11g_{s,\mathrm{re}})^{1/3}a_0 T_0}\right),
	\end{equation}
	with $a_{\mathrm{eq}}$ the equality scale factor, $a_0$ and $T_0=2.7255\,\mathrm{K}$ the present-day values, and $g_{s,\mathrm{re}}$ the entropy degrees of freedom \cite{Dai:2014}. The Hubble parameter at horizon crossing is $H_k = \pi\sqrt{A_s r/2}\,M_{\mathrm{Pl}},$
	with $A_s$ the scalar amplitude and $r$ the tensor-to-scalar ratio.
	
	Tracing the expansion from the pivot scale $k_p$ to the present gives the consistency equation linking inflation to reheating \cite{Liddle:2003as,German:2020iwg,German:2024rmn,German:2025mzg, Cook:2015}:
	\begin{equation}\label{eq:consistency}
		N_k = N_{\mathrm{keq}} - N_{\mathrm{rd}} - N_{\mathrm{re}},
	\end{equation}
	where $N_k\equiv\ln(a_{\mathrm{end}}/a_k)$ is the number of e-folds from horizon crossing to the end of inflation, and $N_{\mathrm{keq}} \equiv \ln\!\left(a_{\mathrm{eq}}H_k/k_p\right)$
	measures the total expansion from horizon exit to equality.
	
	To close the system, $T_{\mathrm{re}}$ must be expressed in terms of $\phi_k$. Decomposing the potential as $V(\phi)=V_0 f(\phi)$ and defining the inflaton mass at the minimum through $m_\phi^2=V''(\phi_0)$ the decay rate depends on the coupling. For gravitational decay into light scalars the rate is $\Gamma_\phi^{(\mathrm{grav})}\approx m_\phi^3/(192\pi M_{\mathrm{Pl}}^2)$, for scalar interactions with Lagrangian $\mathcal{L}_{\mathrm{int}}=-g\phi\chi^2$ it is $\Gamma_\phi^{(g)}\approx g^2/(8\pi m_\phi)$, and for Yukawa couplings with $\mathcal{L}_{\mathrm{int}}=-y\phi\bar{\psi}\psi$ it is $\Gamma_\phi^{(y)}\approx y^2 m_\phi/(8\pi)$ \cite{German:2024rmn, German:2025mzg}. Equating $\rho_{\mathrm{re}}$ to the radiation energy density yields the reheating temperature for each channel,
	\begin{align}
		T_{\mathrm{re}}^{(\mathrm{grav})} &= \frac{90^{1/4}}{\sqrt{192\pi}\,\pi^{1/2}g_{\mathrm{re}}^{1/4}}\;
		\frac{m_\phi^{3/2}}{M_{\mathrm{Pl}}^{1/2}},\\[2mm]
		T_{\mathrm{re}}^{(g)} &= \left(\frac{90}{\pi^2g_{\mathrm{re}}}\right)^{\!1/4}\frac{|\tilde{g}|}{\sqrt{8\pi}}\;
		m_\phi^{-1/2}M_{\mathrm{Pl}}^{3/2},\qquad \tilde{g}\equiv g/M_{\mathrm{Pl}},\\[2mm]
		T_{\mathrm{re}}^{(y)} &= \left(\frac{90}{\pi^2g_{\mathrm{re}}}\right)^{\!1/4}\frac{|y|}{\sqrt{8\pi}}\;
		m_\phi^{1/2}M_{\mathrm{Pl}}^{1/2}.
	\end{align}
	The maximum $T_{\mathrm{re}}$ corresponds to instantaneous reheating ($N_{\mathrm{re}}=0$). Cosmological constraints impose a narrow temperature window, bounded from below by BBN ($T_{\mathrm{re}}\gtrsim10$ MeV) and from above by gravitino overproduction ($T_{\mathrm{re}}\lesssim10^9$ GeV) \cite{Kawasaki:2004qu, Khlopov:1984pf, Ellis:1984eq, Bolz:2000fu, Kohri:2005wn, DeSalas:2015}. The consistency equation ~\eqref{eq:consistency} thus becomes a closed equation for $\phi_k$, solvable numerically once $V(\phi)$ and $\omega_{\mathrm{re}}$ are specified. From the solution for $\phi_k$, all inflationary and reheating quantities ($n_s$, $r$, $N_k$, $N_{\mathrm{re}}$, $T_{\mathrm{re}}$, $m_\phi$, and the couplings $y,\tilde{g}$) follow consistently.

	\section{The \(\alpha\)-Starobinsky model and its minimal extensions}
	
	\subsection{Extension with a multiplicative exponential term}
	
	We propose an extension that preserves the plateau structure by appending a monomial factor $u^{-\delta\sqrt{\alpha}}$ to the potential,
	\begin{equation}\label{eq:V_extended}
		V(u)=V_0(1-u)^2 u^{-\delta\sqrt{\alpha}},\quad u=\exp ~ \!\Bigl(-\sqrt{2/(3\alpha)}\,\phi/M_{\mathrm{Pl}}\Bigr),
	\end{equation}
	where $\alpha>0$ and $\delta\ge0$. This potential reduces to the $\alpha$-Starobinsky model when $\delta=0$. Although the analysis presented here is phenomenological, the form \eqref{eq:V_extended} admits a natural embedding within $\mathcal{N}=1$ supergravity. Mapping to the covering space $\Phi = \phi + i\chi$ via $T = \exp{(\sqrt{2/(3\alpha)}\Phi)}$ and introducing a superpotential that depends on a nilpotent superfield $S$ ($S^2=0$) allows the F-term potential $V = e^K |g(\Phi)|^2$ to be reproduced exactly for a suitable holomorphic function $g(\Phi)$. A complete supergravity construction lies beyond the scope of this work, but the existence of such an embedding indicates that the proposed modification is compatible with supersymmetry and can, in principle, originate from a more fundamental theory \cite{FerraraKalloshLinde2014, KalloshLinde2022, Braglia2023, PallisToumbas2017, Ellis2019, RoestScalisi2015}.
	
	Before analyzing the inflationary dynamics, we verify the stability of the potential. Around the minimum $\phi=0$ ($u=1$), expanding $u=1-\xi$ gives $V(\phi)\approx(V_0/\alpha)\xi^2\propto\phi^2$ for $\alpha>0$, ensuring coherent post-inflationary oscillations. The monotonicity condition in the interval $u\in(0,1)$ is $V_u/V=-2/(1-u)-\delta\sqrt{\alpha}/u<0$, which holds for any $\delta\ge0$. In the large-field limit ($u\to0$), the slow-roll parameter approaches $\epsilon_{\mathrm{plateau}}=\delta^2/3$, so the requirement $\epsilon\ll1$ imposes $0\le\delta\ll1$. Asymptotic stability as $u\to\infty$ yields $V(u)\sim u^{2-\delta\sqrt{\alpha}}$, restricting $\delta<2/\sqrt{\alpha}$, a condition automatically satisfied for small $\delta$ at any $\alpha>0$. All stability criteria are therefore fulfilled within $\alpha>0$ and $0\le\delta<\min\!~\bigl(1,\,2/\sqrt{\alpha}\bigr)$.
	
	\begin{table}[!t]
		\centering
		\footnotesize
		\setlength{\tabcolsep}{3pt}
		\caption{Parameter ranges of the exponential $\alpha$-Starobinsky model that satisfy the $1\sigma$ P-ACT-LB-BK18 constraints, shown for fixed $\alpha=0.5,1.0,5.0$ and several $\omega_{\mathrm{re}}$ intervals.}
		\label{tab:multiplicative_fixed_alpha_delta_e2}
		\begin{tabular}{c c c c c c}
			\toprule
			$\alpha$ & $\omega_{\mathrm{re}}$ & $\delta$ ($10^{-2}$) & $r$ ($10^{-3}$) & $N_k$ & $T_{\mathrm{re}}^{(\mathrm{grav})}$ ($10^9$ GeV) \\
			\midrule
			\multirow{3}{*}{0.5} 
			& $-1/3 < \omega_{\mathrm{re}} \le 0$ & (0.98, 2.93) & (3.2, 8.8) & (44.0, 53.1) & (2.0, 3.9) \\
			& $0 < \omega_{\mathrm{re}} \le 1/3$ & (0.63, 1.69) & (2.4, 4.5) & (53.1, 57.7) & (1.6, 2.4) \\
			& $1/3 < \omega_{\mathrm{re}} < 1$ & (0.28, 1.34) & (1.7, 3.4) & (57.7, 62.4) & (1.2, 2.0) \\
			\midrule
			\multirow{3}{*}{1.0} 
			& $-1/3 < \omega_{\mathrm{re}} \le 0$ & (1.34, 4.17) & (6.1, 17.3) & (44.1, 53.3) & (1.9, 3.6) \\
			& $0 < \omega_{\mathrm{re}} \le 1/3$ & (0.81, 2.40) & (4.4, 8.6) & (53.3, 57.9) & (1.5, 2.3) \\
			& $1/3 < \omega_{\mathrm{re}} < 1$ & (0.28, 1.87) & (3.1, 6.5) & (57.9, 62.7) & (1.2, 1.9) \\
			\midrule
			\multirow{3}{*}{5.0} 
			& $-1/3 < \omega_{\mathrm{re}} \le 0$ & (2.22, 5.41) & (21.9, 38.0) & (47.3, 53.8) & (1.4, 1.9) \\
			& $0 < \omega_{\mathrm{re}} \le 1/3$ & (0.98, 5.41) & (15.2, 37.7) & (53.7, 58.5) & (1.2, 1.7) \\
			& $1/3 < \omega_{\mathrm{re}} < 1$ & (0.10, 4.17) & (10.9, 27.9) & (58.4, 63.4) & (0.9, 1.5) \\
			\midrule
			\multicolumn{2}{c}{Couplings ($\alpha=0.5$)~:} &
			\multicolumn{4}{c}{$y \in (8.4\times10^{-18},\, 8.0\times10^{-7}),\ \tilde{g} \in (1.8\times10^{-22},\, 1.7\times10^{-11})$} \\
			\multicolumn{2}{c}{Couplings ($\alpha=1.0$)~:} & 
			\multicolumn{4}{c}{$y \in (9.8\times10^{-18},\, 9.1\times10^{-7}),\ \tilde{g} \in (1.6\times10^{-22},\, 1.5\times10^{-11})$} \\
			\multicolumn{2}{c}{Couplings ($\alpha=5.0$)~:} & 
			\multicolumn{4}{c}{$y \in (1.2\times10^{-17},\, 1.1\times10^{-6}),\ \tilde{g} \in (1.3\times10^{-22},\, 1.3\times10^{-11})$} \\
			\bottomrule
		\end{tabular}
	\end{table}
	
	Using $u$ as the dynamical variable, the slow-roll parameters follow from equation \eqref{eq:slowroll}. Inflation ends when $\epsilon(u_{\mathrm{end}}) = 1$, which gives
	\begin{equation}\label{eq:u_end_extended}
		u_{\mathrm{end}}(\delta) = \frac{\sqrt{3\alpha} - \delta\sqrt{\alpha}}{2 + \sqrt{3\alpha} - \delta\sqrt{\alpha}} .
	\end{equation}
	
	\begin{figure*}[!t]
		\centering
		\begin{subfigure}[b]{0.33\textwidth}
			\centering
			\includegraphics[width=\textwidth]{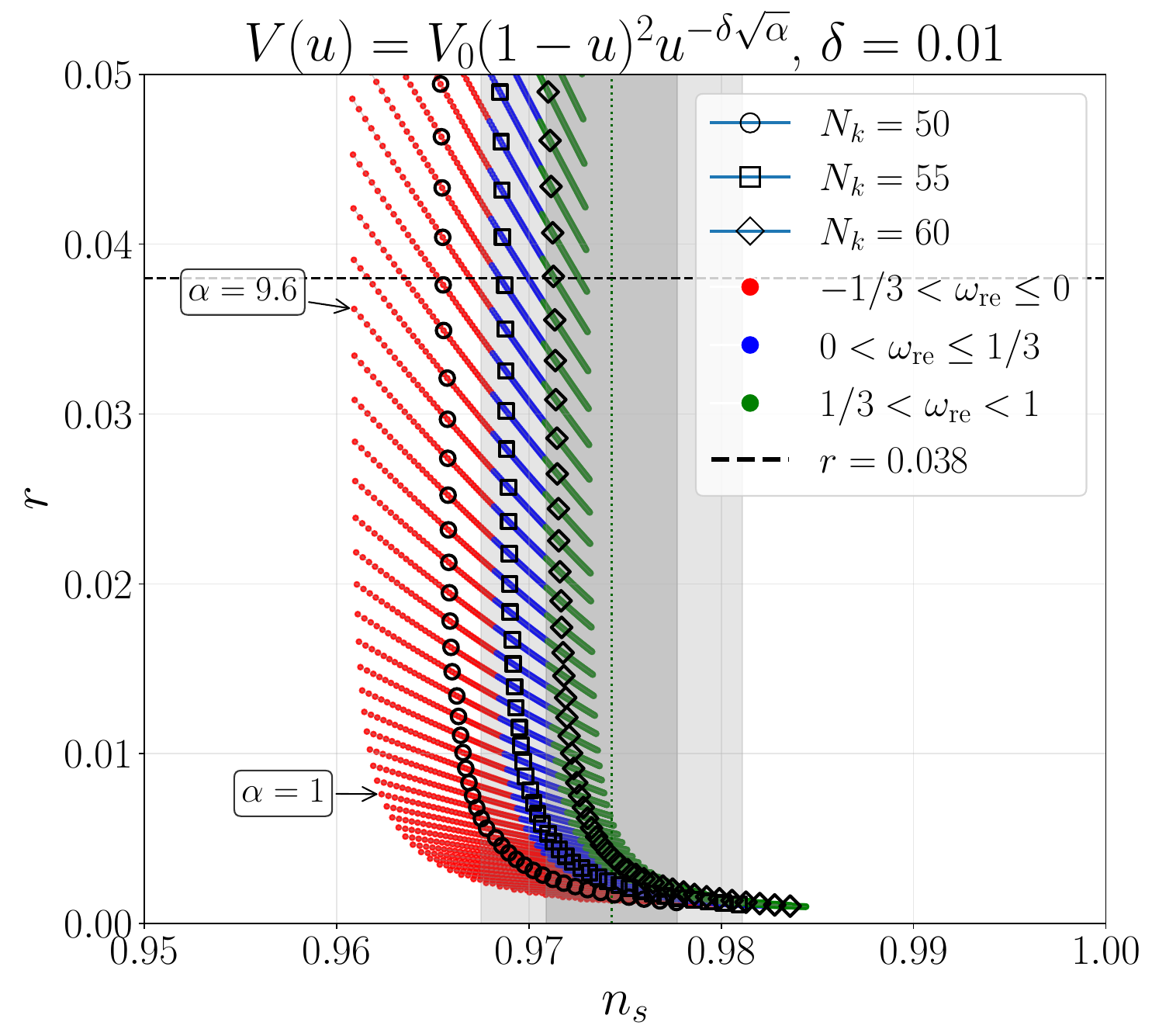}
			\caption{}
			\label{fig:g1}
		\end{subfigure}
		\hfill
		\begin{subfigure}[b]{0.33\textwidth}
			\centering
			\includegraphics[width=\textwidth]{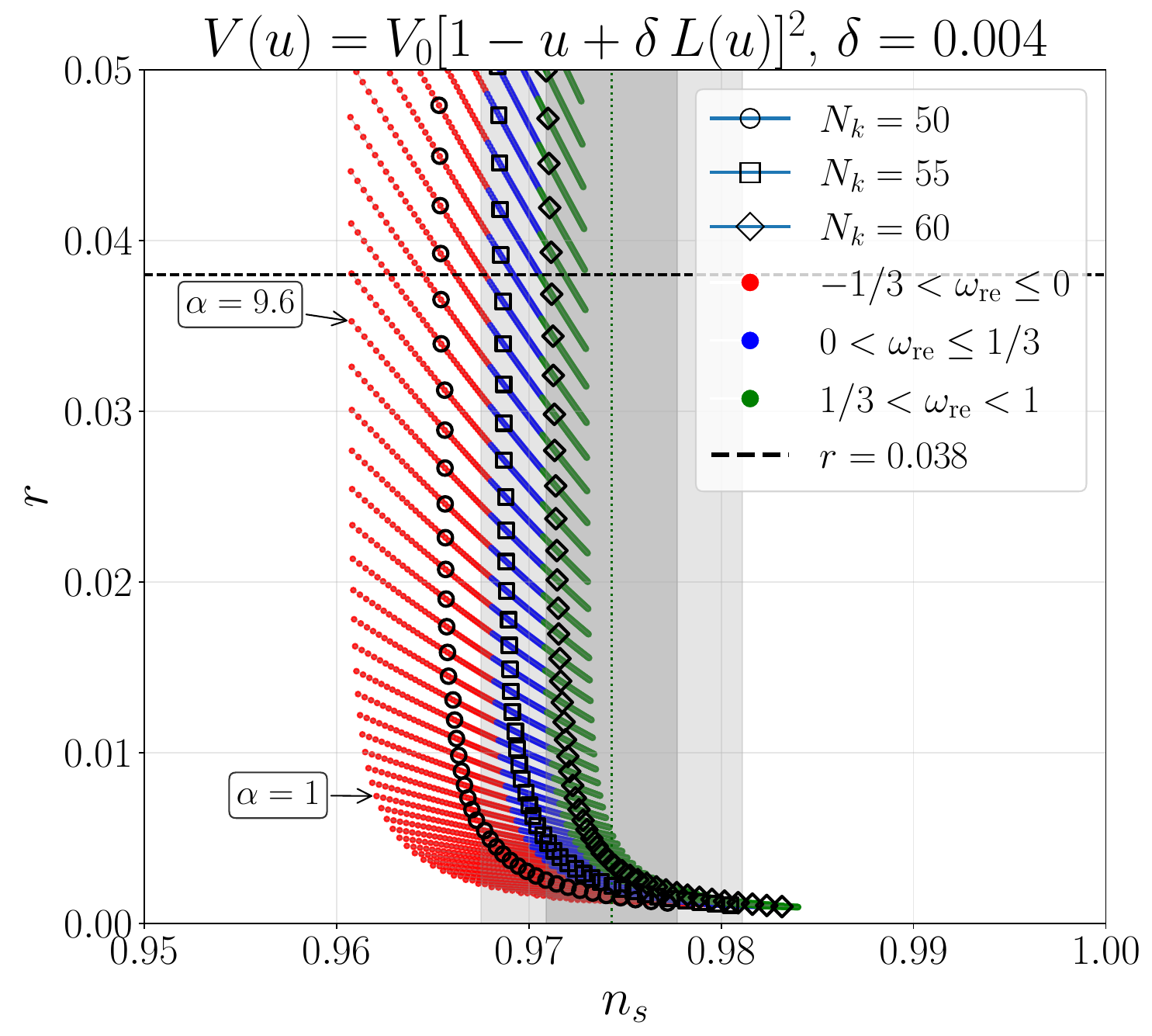}
			\caption{}
			\label{fig:g2}
		\end{subfigure}
		\hfill
		\begin{subfigure}[b]{0.33\textwidth}
			\centering
			\includegraphics[width=\textwidth]{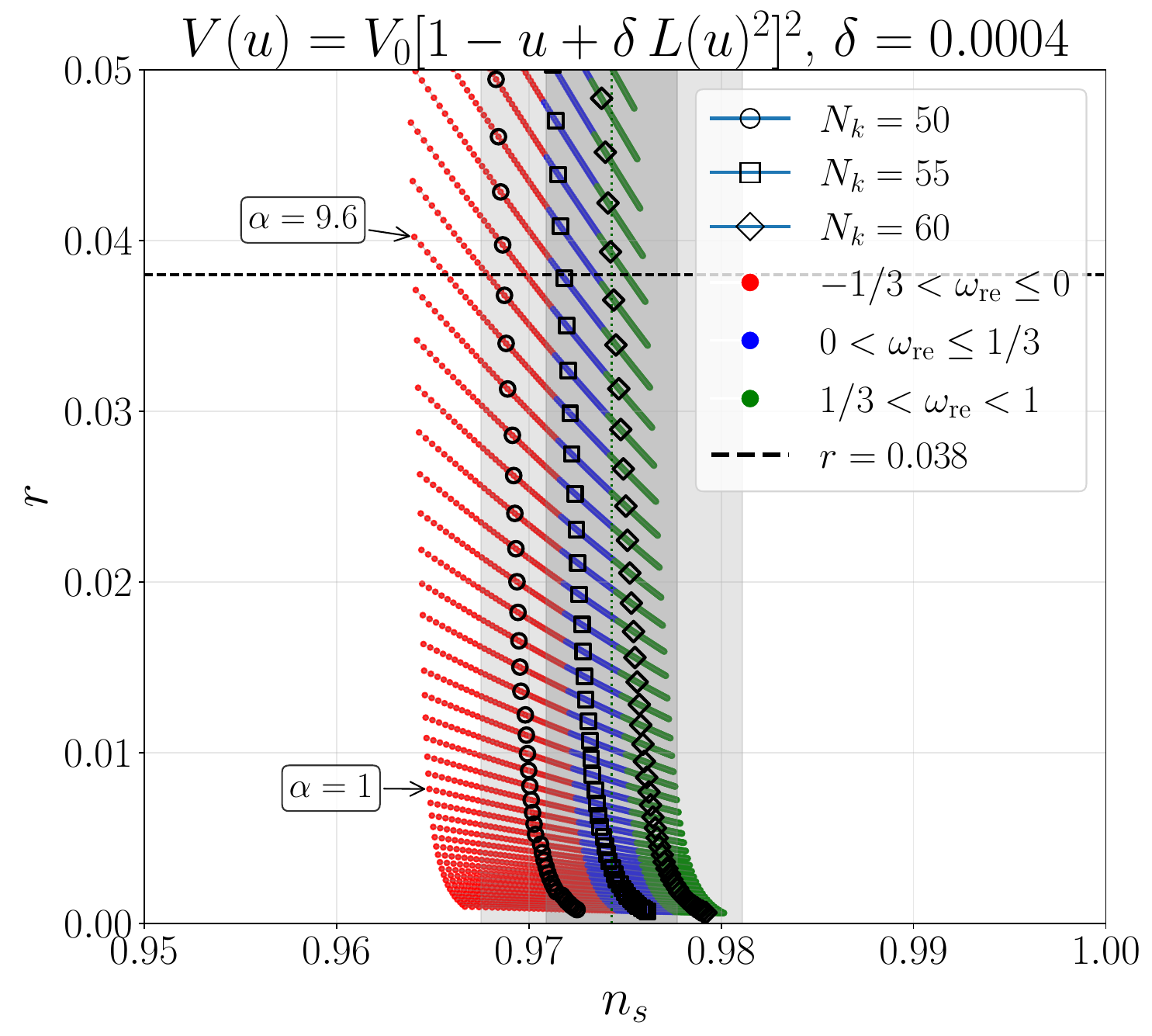}
			\caption{}
			\label{fig:g3}
		\end{subfigure}
		
		\vspace{0.5\baselineskip}
		
		\begin{subfigure}[b]{0.33\textwidth}
			\centering
			\includegraphics[width=\textwidth]{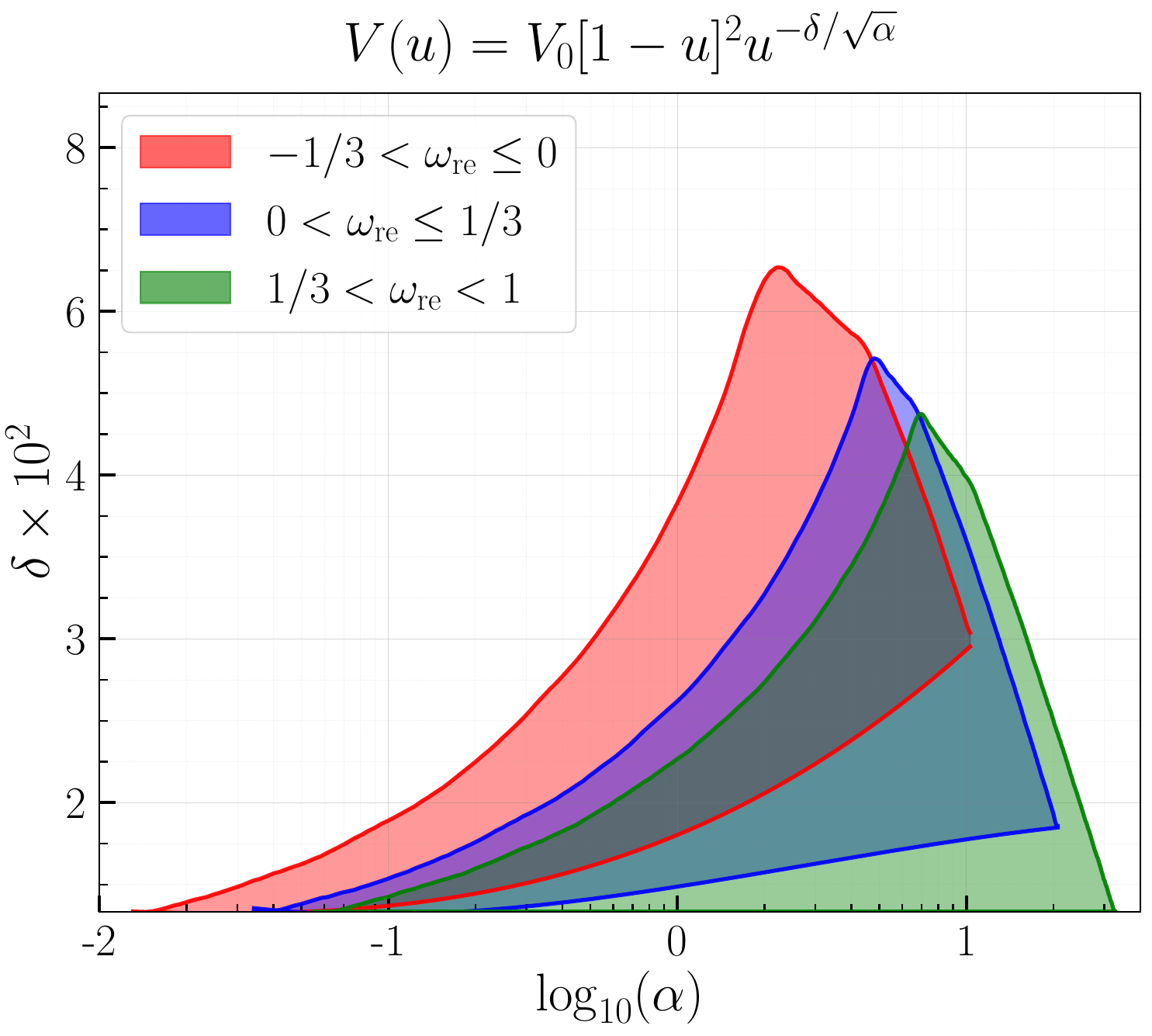}
			\caption{}
			\label{fig:g4}
		\end{subfigure}
		\hfill
		\begin{subfigure}[b]{0.33\textwidth}
			\centering
			\includegraphics[width=\textwidth]{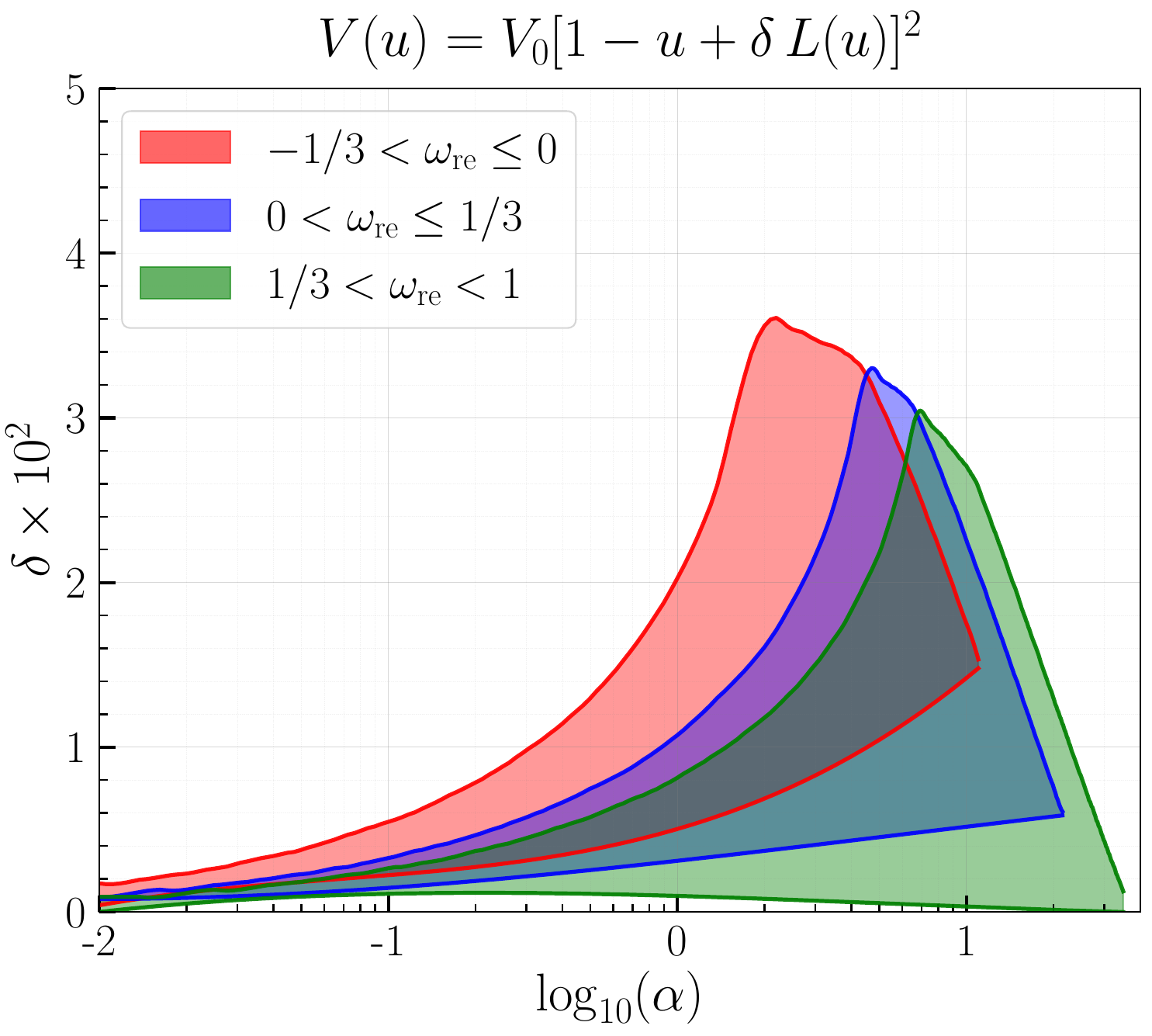}
			\caption{}
			\label{fig:g5}
		\end{subfigure}
		\hfill
		\begin{subfigure}[b]{0.33\textwidth}
			\centering
			\includegraphics[width=\textwidth]{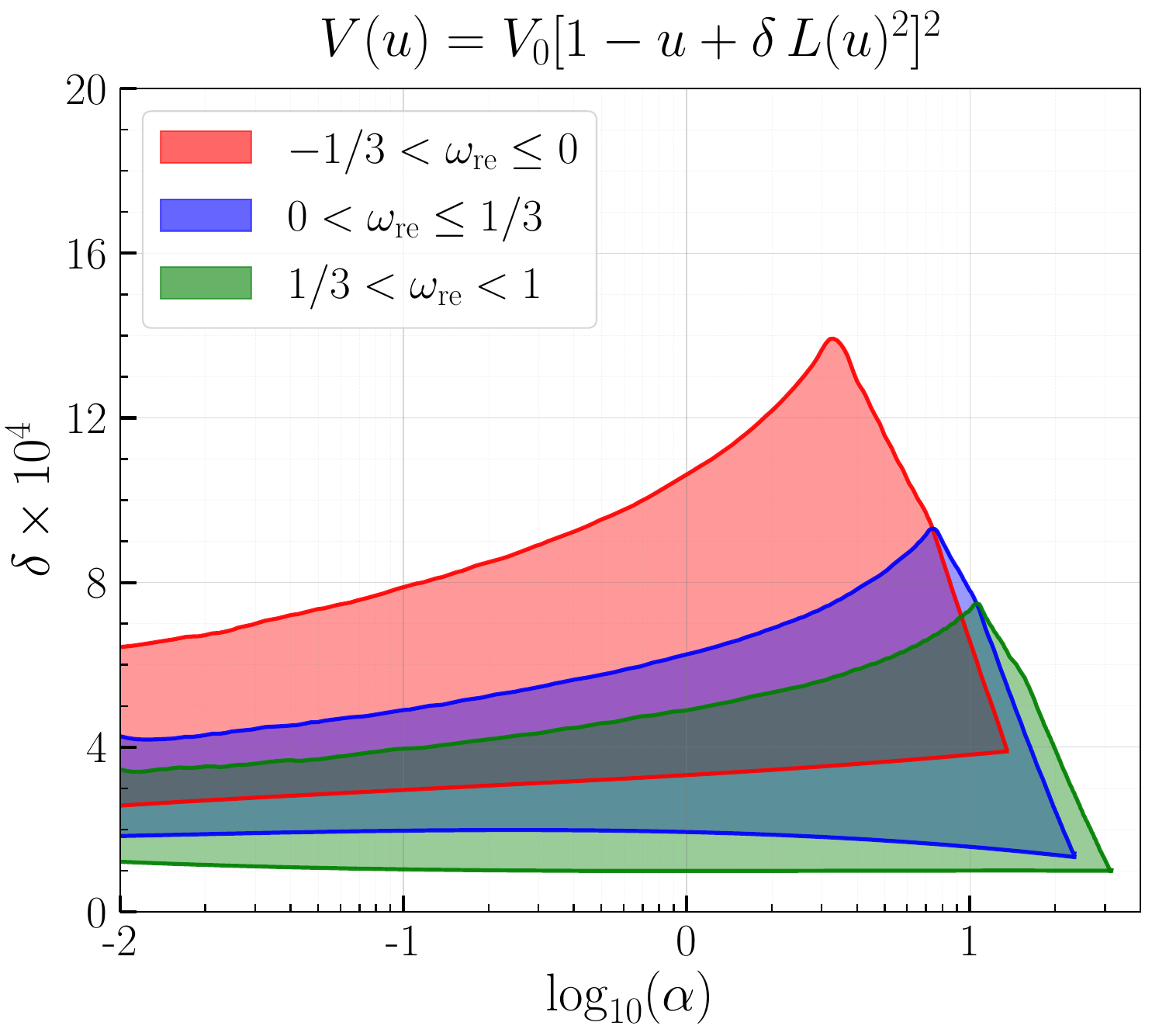}
			\caption{}
			\label{fig:g6}
		\end{subfigure}
		
		\caption{Predictions of $n_s$ and $r$ for the extended $\alpha$-Starobinsky models compared with the P-ACT-LB-BK18 confidence regions. Panels \ref{fig:g1}--\ref{fig:g3} show the $(n_s,r)$ trajectories for the exponential extension ($\delta=0.01$), the polynomial extension with $p=1$ ($\delta=0.004$), and the polynomial extension with $p=2$ ($\delta=0.0004$), respectively. Panels \ref{fig:g4}--\ref{fig:g6} display the corresponding $1\sigma$ allowed parameter space in the $(\alpha,\delta)$ plane for each $\omega_{\mathrm{re}}$ segment.}
		\label{fig:enam_subfigure}
	\end{figure*}
	
	\begin{table*}[!t]
		\centering
		\footnotesize
		\setlength{\tabcolsep}{3pt}
		\caption{Parameter ranges of the additive $\alpha$-Starobinsky model ($p=1$ and $p=2$) satisfying the $1\sigma$ P-ACT-LB-BK18 constraints, for several $\alpha$ and $\omega_{\mathrm{re}}$ intervals.}
		\label{tab:gabungan_aditif}
		\begin{tabular}{c c | c c c c | c c c c}
			\toprule
			\multirow{2}{*}{$\alpha$} & \multirow{2}{*}{$\omega_{\mathrm{re}}$} & \multicolumn{4}{c|}{$p=1$} & \multicolumn{4}{c}{$p=2$} \\
			\cmidrule(lr){3-6} \cmidrule(lr){7-10}
			& & $\delta$ ($10^{-3}$) & $r$ ($10^{-3}$) & $N_k$ & $T_{\mathrm{re}}^{(\mathrm{grav})}$ ($10^9$ GeV) & $\delta$ ($10^{-4}$) & $r$ ($10^{-3}$) & $N_k$ & $T_{\mathrm{re}}^{(\mathrm{grav})}$ ($10^9$ GeV) \\
			\midrule
			\multirow{3}{*}{0.5} 
			& $-1/3 < \omega_{\mathrm{re}} \le 0$ & (3.94, 12.89) & (3.2, 8.4) & (44.0, 53.1) & (2.0, 3.9) & (3.15, 9.26) & (2.9, 6.7) & (43.9, 53.0) & (1.9, 3.4) \\
			& $0 < \omega_{\mathrm{re}} \le 1/3$ & (2.66, 7.78) & (2.5, 4.4) & (53.1, 57.6) & (1.7, 2.5) & (2.08, 5.67) & (2.2, 3.8) & (53.0, 57.6) & (1.5, 2.2) \\
			& $1/3 < \omega_{\mathrm{re}} < 1$ & (1.38, 5.22) & (1.7, 3.2) & (57.7, 62.4) & (1.2, 2.0) & (1.00, 4.59) & (1.7, 2.9) & (57.6, 62.4) & (1.2, 1.8) \\
			\midrule
			\multirow{3}{*}{1.0} 
			& $-1/3 < \omega_{\mathrm{re}} \le 0$ & (5.22, 19.29) & (5.8, 18.0) & (44.1, 53.3) & (1.8, 3.8) & (3.51, 10.33) & (5.3, 12.4) & (44.0, 53.2) & (1.7, 3.2) \\
			& $0 < \omega_{\mathrm{re}} \le 1/3$ & (3.94, 10.34) & (4.3, 9.0) & (53.3, 57.9) & (1.5, 2.4) & (2.08, 6.38) & (4.0, 7.0) & (53.2, 57.8) & (1.4, 2.1) \\
			& $1/3 < \omega_{\mathrm{re}} < 1$ & (1.38, 7.78) & (3.4, 6.5) & (57.9, 62.7) & (1.2, 1.9) & (1.00, 4.95) & (3.1, 5.5) & (57.9, 62.6) & (1.2, 1.8) \\
			\midrule
			\multirow{3}{*}{5.0} 
			& $-1/3 < \omega_{\mathrm{re}} \le 0$ & (10.34, 30.81) & (21.5, 37.9) & (47.0, 53.8) & (1.5, 2.0) & (3.51, 11.41) & (19.6, 37.9) & (44.2, 53.7) & (1.4, 2.2) \\
			& $0 < \omega_{\mathrm{re}} \le 1/3$ & (5.22, 32.09) & (15.6, 38.0) & (53.7, 58.5) & (1.2, 1.8) & (1.72, 8.54) & (14.7, 27.7) & (53.7, 58.4) & (1.2, 1.7) \\
			& $1/3 < \omega_{\mathrm{re}} < 1$ & (1.38, 21.85) & (11.3, 28.0) & (58.4, 63.4) & (1.0, 1.6) & (1.00, 6.03) & (11.8, 21.1) & (58.4, 63.3) & (1.0, 1.5) \\
			\midrule
			\multicolumn{2}{c|}{Couplings ($\alpha=0.5$)} & 
			\multicolumn{4}{c|}{$y\in(1.0\times10^{-17},\,9.5\times10^{-7}),\ \tilde{g}\in(1.5\times10^{-22},\,1.4\times10^{-11})$} & 
			\multicolumn{4}{c}{$y\in(8.3\times10^{-18},\,8.1\times10^{-7}),\ \tilde{g}\in(1.8\times10^{-22},\,1.7\times10^{-11})$} \\
			\multicolumn{2}{c|}{Couplings ($\alpha=1.0$)} & 
			\multicolumn{4}{c|}{$y\in(1.1\times10^{-17},\,1.0\times10^{-6}),\ \tilde{g}\in(1.4\times10^{-22},\,1.4\times10^{-11})$} & 
			\multicolumn{4}{c}{$y\in(1.1\times10^{-17},\,1.0\times10^{-6}),\ \tilde{g}\in(1.4\times10^{-22},\,1.3\times10^{-11})$} \\
			\multicolumn{2}{c|}{Couplings ($\alpha=5.0$)} & 
			\multicolumn{4}{c|}{$y\in(1.1\times10^{-17},\,1.1\times10^{-6}),\ \tilde{g}\in(1.4\times10^{-22},\,1.3\times10^{-11})$} & 
			\multicolumn{4}{c}{$y\in(1.1\times10^{-17},\,1.1\times10^{-6}),\ \tilde{g}\in(1.4\times10^{-22},\,1.3\times10^{-11})$} \\
			\bottomrule
		\end{tabular}
	\end{table*}
	
	Defining $\beta := \delta\sqrt{\alpha}$, the number of e-folds takes the exact form
	\begin{equation}\label{eq:Nk_extended}
		N_k(u_k) = \frac{3\sqrt{\alpha}}{\beta} \left[ 
		\frac{1}{2}\ln\frac{u_{\mathrm{end}}}{u_k} - 
		\frac{1}{2-\beta} \ln\!\left( 
		\frac{\beta + (2-\beta) u_{\mathrm{end}}}
		{\beta + (2-\beta) u_k} \right) \right]. 
	\end{equation}
	The inversion $u_k(N_k)$ can be obtained perturbatively in the limit $\beta \ll 1$. Up to order $\beta^2$, it gives
	\begin{equation}
		u_k(N_k,\beta) \approx \frac{3\alpha}{4N_k} - \frac{\beta}{4} + \beta^2\left[ \frac{1}{12} - \frac{1}{6N_k}\ln\!\left(\frac{4N_k}{3\alpha}\right) \right].
		\label{eq:uk_analitik2_beta}
	\end{equation}
	To express the observables directly in terms of $N_k$, we take the limit $N_k \gg 1$ and expand $u_k(N_k)$ to first order, $u_k \approx 3\alpha/(4N_k) - \beta/4$. Defining the deviations from the standard $\alpha$-Starobinsky predictions as $\Delta n_s:=n_s(N_k)-n_s^{(0)}$ and similarly for the other observables, we find
	\begin{align}
		\Delta n_s \approx \frac{4\beta}{N_k} + \frac{2\beta}{3\alpha} - \frac{\beta^2}{2\alpha}, \quad 
		\Delta r \approx \frac{16\beta^2}{3\alpha}, \quad
		\Delta n_{sk} \approx  - \frac{4\beta}{N_k^2}. \label{eq:nsk_extended_beta}
	\end{align}
	From equation \eqref{eq:nsk_extended_beta}, all corrections vanish for $\beta=0$, whereas $\beta>0$ contributes a positive linear shift to $n_s$ that places it within the range favored by P-ACT-LB-BK18. The quadratic correction to $r$, of order $\beta^2/\alpha$, remains negligible for $\beta\lesssim0.02\sqrt{\alpha}$, while $\Delta n_{sk}\propto 1/N_k^2$ stays small and negative as long as $N_k\gg 1$.
	
	The numerical results obtained by solving the consistency relation \eqref{eq:consistency} are presented in Figs.~\ref{fig:g1} and~\ref{fig:g4}. Figure~\ref{fig:g1} displays the $(n_s,r)$ predictions for a fixed representative value of $\delta$, while Fig.~\ref{fig:g4} illustrates the $1\sigma$ allowed parameter space in the $(\alpha,\delta)$ plane for each $\omega_{\mathrm{re}}$ segment. Representative values for selected $\alpha$ are collected in Table~\ref{tab:multiplicative_fixed_alpha_delta_e2}. In contrast to the original $\alpha$-Starobinsky model shown in Fig.~\ref{fig:alpha_starobinsky}, the exponential deformation clearly improves the agreement with the observational constraints. From Fig.~\ref{fig:g4}, one can infer that the viable $1\sigma$ parameter space requires $\alpha \lesssim 35$ with $\delta\sim\mathcal{O}(10^{-2})$. For $\alpha=1$, the extended model covers the $1\sigma$ P-ACT-LB-BK18 band with $\delta\in(0.0028,0.0417)$, $r\in(0.0031,0.0173)$, and $N_k\in(44.1,62.7)$, reaching $n_s\approx0.9743$ at $\delta\approx0.0169$. The associated gravitational reheating temperature ranges $(1.2\text{--}3.6)\times10^9$ GeV and safely satisfies both the BBN and gravitino constraints. The decay couplings compatible with these reheating bounds are $9.8\times10^{-18}\le y\le9.1\times10^{-7}$ for the Yukawa channel and $1.6\times10^{-22}\le \tilde{g}\le1.5\times10^{-11}$ for the scalar coupling. Additional results for $\alpha=0.5$ and $\alpha=5.0$ are listed in Table~\ref{tab:multiplicative_fixed_alpha_delta_e2}. Requiring the usual e-folding number $N_k\in[50,65]$ further restricts the reheating equation of state to $0<\omega_{\mathrm{re}}\le1$.
	\subsection{Extension with an additive polynomial term}
	
	The second extension appends a polynomial additive term to the $\alpha$-Starobinsky potential:
	\begin{equation}\label{eq:V_additive}
		V(u)=V_0\bigl[1-u+\delta\,L(u)^p\bigr]^2,\qquad 
		L(u)\equiv-\sqrt{\frac{3\alpha}{2}}\ln u,
	\end{equation}
	with $\alpha>0$, $p\in\mathbb{Z}^+$, and $\delta\ge0$. As with the exponential case, this additive potential admits an embedding into $\mathcal{N}=1$ supergravity through covering space constructions and nilpotent superfields, following standard $\alpha$-attractor techniques \cite{Braglia2023, FerraraKalloshLinde2014, RoestScalisi2015}; a detailed analysis is again beyond the present scope.
	
	The stability conditions restrict the viable parameter space. Around the minimum $\phi=0$ ($u=1$), the expansion $L(u)\approx\sqrt{3\alpha/2}(1-u)$ suppresses the perturbative term by $(1-u)^p$, preserving the harmonic form $V\propto\phi^2$ and ensuring coherent post-inflationary oscillations. Monotonicity along the inflationary trajectory requires $df/du<0$ for $f(u)=1-u+\delta(-\sqrt{3\alpha/2}\ln u)^p$. The derivative $df/du=-1-\frac{\delta p}{u}\sqrt{3\alpha/2}L(u)^{p-1}$ is always negative for $\delta\ge0$; a negative $\delta$ would induce a false vacuum that halts inflation prematurely. Perturbativity at horizon crossing ($u_k\approx3\alpha/(4N_k)$) demands $|\delta L(u)^p|/(1-u)\ll1$, which translates to $\delta\ll\bigl[\frac{3\alpha}{2}\bigr]^{-p/2}\bigl[\ln\frac{8N_k}{9\alpha}\bigr]^{-p}$. This bound is readily satisfied for $p=1,2$ with sufficiently small $\delta$, but diverges for $p\ge3$ even at $\delta\sim10^{-5}$. The physical parameter space is therefore effectively restricted to $p\in\{1,2\}$.

	The slow-roll parameters follow from equation \eqref{eq:slowroll}, but the condition $\epsilon(u_{\text{end}}) = 1$ is not analytically tractable in closed form. Within the perturbative framework, the end of inflation is approximated by
	\begin{equation}\label{eq:u_end_additive}
		u_{\text{end}} \simeq  \frac{\sqrt{3\alpha}}{\sqrt{3\alpha}+2} + \delta\, \frac{L(u_0)^{p-1}}{1 + 2/\sqrt{3\alpha}} \left[ L(u_0) - \sqrt{2}p \right]. 
	\end{equation}
	The number of e-folds $N(u_k)$ is computed perturbatively in $\delta$ using the exact integral
	\begin{equation}\label{eq:I_nm_def}
		\begin{split}
			\mathcal{I}_{n,m}(u) &\equiv \int^u x^{-n} L(x)^m dx \\
			&= -\frac{u^{1-n}}{n-1} \sum_{j=0}^m \frac{m!}{(m-j)!} \frac{L(u)^{m-j}}{\bigl[(n-1)\sqrt{2/(3\alpha)}\bigr]^j}, 
		\end{split}
	\end{equation}
	valid for $n > 1$. Expanding $N(u_k)$ to order $\delta^2$ yields
	\begin{align}
		N^{(2)}(u_k) &= \bigl[\mathcal{H}_0(u_{\text{end}}) - \mathcal{H}_0(u_k)\bigr] + \delta \bigl[\mathcal{F}_1(u_{\text{end}};p) - \mathcal{F}_1(u_k;p)\bigr] \nonumber\\
		&\quad + \delta^2 \bigl[\mathcal{F}_2(u_{\text{end}};p) - \mathcal{F}_2(u_k;p)\bigr], \label{eq:N_additive}
	\end{align}
	where $\mathcal{H}_0$, $\mathcal{F}_1$, and $\mathcal{F}_2$ are linear combinations of $\mathcal{I}_{n,m}$:
	\begin{align}
		\mathcal{H}_0(u) &= -\frac{\Gamma^2}{2}\left(\frac{1}{u}+\ln u\right), \label{eq:H0_add} \\
		\mathcal{F}_1(u;p) &= \frac{\Gamma^2}{2} \mathcal{I}_{2,p}(u) 
		+ \frac{p}{2}\Gamma^3 \bigl( \mathcal{I}_{2,p-1}(u) - \mathcal{I}_{3,p-1}(u) \bigr), \label{eq:F1_add} \\
		\mathcal{F}_2(u;p) &= \frac{p^2}{2}\Gamma^4 \bigl( \mathcal{I}_{4,2p-2}(u) - \mathcal{I}_{3,2p-2}(u) \bigr)
		- \frac{p}{2}\Gamma^3 \mathcal{I}_{3,2p-1}(u). \label{eq:F2_add}
	\end{align}
	With the definition \eqref{eq:I_nm_def}, the integrals $\mathcal{I}_{n,m}(u)$ evaluate exactly for any $p \in \mathbb{Z}^+$, reducing $\mathcal{F}_1$ and $\mathcal{F}_2$ to polynomial-logarithmic forms. Introducing $\Gamma := \sqrt{3\alpha/2}$, the explicit expressions for $p=1,2$ are
	\begin{align}
		\mathcal{F}_1(u;1) &= -\frac{\Gamma^2 L(u)}{2u} + \Gamma^3\left(\frac{1}{4u^2}-\frac{1}{u}\right), \\
		\mathcal{F}_2(u;1) &= \frac{\Gamma^2 L(u)}{2u^2} + \Gamma^4\left(\frac{3}{8u^2}-\frac{1}{6u^3}\right), \\
		\mathcal{F}_1(u;2) &= -\frac{\Gamma^2 L(u)^2}{2u} + \frac{\Gamma^3 L(u)}{2u^2}(1-4u) + \frac{\Gamma^4}{4u^2}(1-8u), \\
		\mathcal{F}_2(u;2) &= \frac{\Gamma^3 L(u)^3}{2u^2} + \frac{\Gamma^4 L(u)^2}{12u^3}(21u-8) \nonumber\\
		&\quad + \frac{\Gamma^5 L(u)}{36u^3}(63u-16) + \frac{\Gamma^6}{216u^3}(189u-32).
	\end{align}
	
	To express the observables directly as functions of $N_k$, we invert equation \eqref{eq:N_additive} analytically using the Lambert $W_{-1}$ function. The zeroth-order solution satisfies $u_k^{-1} + \ln u_k = \mathcal{C}(N_k)$ with $\mathcal{C}(N_k) = \frac{3\alpha}{2} N_k + u_0^{-1} + \ln u_0$. Substituting $x = u_k^{-1}$ reduces this to the canonical form $-x e^{-x} = -e^{-\mathcal{C}}$, giving the exact solution
	\begin{equation}\label{eq:uk0_lambert_add}
		u_k^{(0)}(N_k) = -\frac{1}{W_{-1}\!\bigl(-e^{-\mathcal{C}(N_k)}\bigr)}.
	\end{equation}
	Perturbative corrections up to order $\delta^2$ follow from implicit differentiation,
	\begin{equation}\label{eq:uk_perturb_add}
		u_k^{(2)}(N_k) \simeq  u_k^{(0)} + \delta\, u_k^{(1)} + \delta^2\, u_k^{(2)} ,
	\end{equation}
	with the coefficients expressed through derivatives of \eqref{eq:H0_add}--\eqref{eq:F2_add}:
	\begin{align}
		u_k^{(1)}(N_k) &= -\bigl(\mathcal{F}_1\!\bigl(u_k^{(0)}; p\bigr) - \mathcal{F}_1\!\bigl(u_{\text{end},0}; p\bigr) \bigr) / \mathcal{H}_0'\!\bigl(u_k^{(0)}\bigr), \label{eq:uk1_explicit} \\
		u_k^{(2)}(N_k) &= -\frac{1}{\mathcal{H}_0'\!\bigl(u_k^{(0)}\bigr)} \left[ \frac{1}{2}\mathcal{H}_0''\!\bigl(u_k^{(0)}\bigr)\bigl(u_k^{(1)}\bigr)^2 \right. \nonumber\\
		&\quad \left. + \mathcal{F}_1'\!\bigl(u_k^{(0)}; p\bigr) u_k^{(1)} + \mathcal{F}_2\!\bigl(u_k^{(0)}; p\bigr) - \mathcal{F}_2\!\bigl(u_{\text{end},0}; p\bigr) \right]. \label{eq:uk2_explicit}
	\end{align}
	The required derivatives of $\mathcal{H}_0$ are $\mathcal{H}_0'(u) = \frac{3\alpha}{2}\frac{1-u}{u^2}$ and $\mathcal{H}_0''(u) = \frac{3\alpha}{2}\frac{u-2}{u^3}$. All components in \eqref{eq:uk1_explicit} and \eqref{eq:uk2_explicit} are available in closed analytical form, enabling a complete computation of $u_k(N_k)$.

	Substituting the perturbative series $u_k(N_k) \simeq u^{(0)} + \delta u^{(1)} + \delta^2 u^{(2)}$ into the slow-roll parameters and Taylor-expanding to second order in $\delta$ yields the observables as explicit functions of $N_k$, $\delta$, and the potential parameters. Defining the auxiliary variable $\Lambda := \ln\!\left(4N_k/(3\alpha)\right)$, we obtain the following approximations.
	
	\noindent
	For $p=1$:
	\begin{align}
		\Delta n_s & \simeq 
		(\delta\,\Gamma) \Bigl[ \frac{4\Lambda - 8}{N_k} + \frac{6\alpha\Lambda}{N_k^2} \Bigr] + (\delta \Gamma)^2 \Bigl[ \frac{2(2\Lambda^2 - 8\Lambda + 6)}{N_k}\nonumber\\
		&\qquad \qquad  + \frac{3\alpha(2\Lambda^2 - 3\Lambda)}{N_k^2} - \frac{9\alpha^2}{4\Gamma^2 N_k^3} \Bigr], \label{eq:ns_p1_Nk_ord2} \\
		\Delta r &\simeq  (\delta\,\Gamma) \Bigl[ \frac{32}{N_k} - \frac{24\alpha\Lambda}{N_k^2} \Bigr] + (\delta\,\Gamma)^2 \Bigl[ \frac{16(2\Lambda - 3)}{N_k} \nonumber\\
		&\qquad \qquad - \frac{24\alpha(2\Lambda - 1)}{N_k^2} + \frac{36\alpha^2}{\Gamma^2 N_k^3} \Bigr], \label{eq:r_p1_Nk_ord2} \\
		\Delta n_{sk} &\simeq 
		(\delta\,\Gamma) \Bigl[ \frac{4 - 4\Lambda}{N_k^2} + \frac{12\alpha(\Lambda-1)}{N_k^3} \Bigr] + (\delta\,\Gamma)^2 \Bigl[ \frac{4(\Lambda - \Lambda^2)}{N_k^2} \nonumber\\
		&\qquad \qquad  + \frac{12\alpha(2\Lambda - \Lambda^2 - 1)}{N_k^3} + \frac{27\alpha^2}{2\Gamma^2 N_k^4} \Bigr]. \label{eq:nsk_p1_Nk_ord2}
	\end{align}
	
	For $p=2$:
	\begin{align}
		\Delta n_s &\simeq (\delta\,\Gamma^2) \Bigl[ \frac{8\Lambda(\Lambda - 2)}{N_k} + \frac{12\alpha\Lambda^2}{N_k^2} \Bigr] + (\delta\,\Gamma^2)^2 \Bigl[\frac{6\alpha(3\Lambda^3 - 4\Lambda^2)}{N_k^2}   \nonumber\\
		&\qquad \qquad +  \frac{4(3\Lambda^3 - 12\Lambda^2 + 12\Lambda - 4)}{N_k}  - \frac{9\alpha^2\Lambda^2}{2\Gamma^4 N_k^3} \Bigr], \label{eq:ns_p2_Nk_ord2} \\
		\Delta r &\simeq  (\delta\,\Gamma^2) \Bigl[ \frac{64\Lambda}{N_k} - \frac{48\alpha\Lambda^2}{N_k^2} \Bigr] + (\delta\,\Gamma^2)^2 \Bigl[ \frac{32(3\Lambda^2 - 2\Lambda)}{N_k} \nonumber\\
		&\qquad \qquad  - \frac{48\alpha(3\Lambda^2 - \Lambda)}{N_k^2} + \frac{72\alpha^2\Lambda^2}{\Gamma^4 N_k^3} \Bigr], \label{eq:r_p2_Nk_ord2} \\[2mm]
		\Delta n_{sk}&\simeq 
		(\delta\,\Gamma^2) \Bigl[ \frac{8\Lambda(1-\Lambda)}{N_k^2} + \frac{24\alpha\Lambda(\Lambda-1)}{N_k^3} \Bigr] + (\delta\,\Gamma^2)^2 \Bigl[ \frac{8\Lambda^2(1-\Lambda)}{N_k^2} \nonumber\\
		&\qquad \qquad  + \frac{24\alpha\Lambda^2(2\Lambda-3)}{N_k^3} + \frac{54\alpha^2\Lambda^2}{\Gamma^4 N_k^4} \Bigr]. \label{eq:nsk_p2_Nk_ord2}
	\end{align}
	The expansions \eqref{eq:ns_p1_Nk_ord2}--\eqref{eq:nsk_p2_Nk_ord2} reduce consistently to the $\alpha$-Starobinsky limit when $\delta=0$. The corrections at order $\delta^1$ and $\delta^2$ exhibit polynomial-logarithmic growth with $p$, scaling as $\propto\Lambda^{p-1}$ and $\propto\Lambda^{2p-2}$ respectively. For $\delta>0$, both $\Delta n_s$ and $\Delta r$ remain positive throughout $N_k\gg 1$, shifting $n_s$ above $n_s^{(0)}$ into better agreement with P-ACT-LB-BK18 while respecting $r<0.038$. The $\delta^2$ contribution to $n_s$ is subdominant compared to the linear term for $\delta\ll1$, but becomes relevant for precision analyses.

	The numerical results summarized in Table~\ref{tab:gabungan_aditif} and illustrated in Figs.~\ref{fig:g2}--\ref{fig:g3} and Figs.~\ref{fig:g5}--\ref{fig:g6} demonstrate that both additive models with $p=1,2$ consistently satisfy the P-ACT-LB-BK18 constraints. From the parameter space plots in Figs.~\ref{fig:g5} and~\ref{fig:g6}, the viable $1\sigma$ region extends to $\alpha \lesssim 35$, while the representative values collected in the table focus on the $\alpha\lesssim1$ case. Increasing $\omega_{\mathrm{re}}$ systematically lowers $\delta$ and $T_{\mathrm{re}}$ while raising $N_k$, and within the conventional range $N_k\in[50,65]$ the viable equation of state is found to be $0<\omega_{\mathrm{re}}\le1$. For the $p=1$ model, one requires $\delta\sim\mathcal{O}(10^{-3})$, for example $(3.9\text{--}10.3)\times10^{-3}$ at $\alpha=1$, with $T_{\mathrm{re}}\sim(1.5\text{--}2.4)\times10^9$ GeV. In contrast, the perturbatively more natural $p=2$ model needs only $\delta\sim\mathcal{O}(10^{-4})$, for instance $(2.3\text{--}5.9)\times10^{-4}$, to achieve the same spectral predictions, with $T_{\mathrm{re}}\sim(1.4\text{--}2.1)\times10^9$ GeV. The allowed parameter space confirms that $p=1$ requires $\delta\sim\mathcal{O}(10^{-2})$ while $p=2$ only needs $\delta\sim\mathcal{O}(10^{-4})$ to meet the constraints. All viable scenarios yield reheating temperatures safe from BBN and gravitino bounds, and the corresponding decay couplings are $y\lesssim10^{-6}$ and $\tilde{g}\lesssim10^{-11}$. Although $p=1$ offers a larger detectable signal, $p=2$ is preferred by perturbative stability.
	
	\section{Conclusions}
	\label{sec:conclusion}
	
	We have explored two minimal extensions of the \(\alpha\)-Starobinsky inflation model, a multiplicative exponential modification and an additive polynomial deformation, in response to the observational tension induced by the latest ACT DR6 data. Through an analytical treatment up to second order in the perturbative parameter \(\delta\) and a full numerical evaluation incorporating reheating, we have shown that both models shift the scalar spectral index \(n_s\) into the \(1\sigma\) confidence region of the joint P-ACT-LB-BK18 dataset while respecting the strict bound \(r < 0.038\).

	The central result of this work is the consistent integration of reheating dynamics via the consistency equation. The parameter space analysis from the plots shows that the $1\sigma$ region extends to $\alpha \lesssim 35$. When the usual e-folding range $N_k \in [50, 65]$ is imposed, the viable reheating equation of state satisfies $0<\omega_{\mathrm{re}}\le1$. Among the scenarios considered, the additive model with $p=2$ emerges as the most natural from a perturbative perspective, as it requires only $\delta \sim \mathcal{O}(10^{-4})$ and yields a reheating temperature $T_{\mathrm{re}} \sim 10^9$ GeV that safely lies between the BBN lower bound and the gravitino overproduction limit. The exponential model, by contrast, requires a somewhat larger correction, $\delta \sim \mathcal{O}(10^{-2})$, which compensates by producing a spectral deviation that may be more accessible to next-generation CMB experiments. Both classes of models preserve the plateau structure and attractor properties of the original $\alpha$-Starobinsky framework, thereby ensuring their theoretical consistency while extending their phenomenological reach to accommodate the latest precision cosmological data.
	
	\bibliographystyle{elsarticle-num}
	\bibliography{refs2}
	
\end{document}